\documentclass[12pt]{article}
\makeatletter

\usepackage{graphicx}%
\usepackage{multirow}%
\usepackage{amsmath,amssymb,amsfonts}%
\usepackage{float}%
\usepackage{amsthm}%
\usepackage{mathrsfs}%
\usepackage{xcolor}%
\usepackage{textcomp}%
\usepackage{manyfoot}%
\usepackage{url}%

\usepackage{booktabs}%
\usepackage{algorithm}%

\usepackage{algorithmicx}%
\usepackage{algpseudocode}%
\usepackage{listings}%
\usepackage{tabularx}  
\usepackage{booktabs}
\usepackage{braket} 

\theoremstyle{thmstyleone}%

\theoremstyle{thmstyletwo}%

\theoremstyle{thmstylethree}%

\def\BibTeX{{\rm B\kern-.05em{\sc i\kern-.025em b}\kern-.08em
    T\kern-.1667em\lower.7ex\hbox{E}\kern-.125emX}}

\usepackage{cite}
\usepackage{booktabs} 

\usepackage[top=0.75in,bottom=1in,left=0.75in,right=0.75in,columnsep=0.2in]{geometry} 

\usepackage{blindtext}
\usepackage{eso-pic}

\usepackage{eso-pic}
\usepackage{graphicx}
\def\BibTeX{{\rm B\kern-.05em{\sc i\kern-.025em b}\kern-.08em
    T\kern-.1667em\lower.7ex\hbox{E}\kern-.125emX}}

\begin{document}

\title{A Quantum-Secure Voting Framework Using QKD, Dual-Key Symmetric Encryption, and Verifiable Receipts}
\author{Taha M. Mahmoud \\ University of North Dakota \\ \texttt{taha.mahmoud@und.edu} 
\and Naima Kaabouch \\ University of North Dakota \\ \texttt{naima.kaabouch@und.edu}}
\date{}

\maketitle
\begin{center}
\textbf{Disclaimer:} This is the author's accepted manuscript of the paper published in \textit{IEEE International Conference on Artificial Intelligence, Computer, Data Sciences and Applications (ACDSA 2025)}. 
The published version is available at IEEE Xplore: 
\url{https://doi.org/10.1109/ACDSA65407.2025.11165862}
\end{center}

\begin{abstract}
Electronic voting systems face growing risks from cyberattacks and data breaches, which are expected to intensify with the advent of quantum computing. To address these challenges, we introduce a quantum-secure voting framework that integrates Quantum Key Distribution (QKD), Dual-Key Symmetric Encryption, and verifiable receipt mechanisms to strengthen the privacy, integrity, and reliability of the voting process. The framework enables voters to establish encryption keys securely, cast encrypted ballots, and verify their votes through receipt-based confirmation, all without exposing the vote contents. To evaluate performance, we simulate both quantum and classical communication channels using the Message Queuing Telemetry Transport (MQTT) protocol. Results demonstrate that the system can process large numbers of votes efficiently with low latency and minimal error rates. This approach offers a scalable and practical path toward secure, transparent, and verifiable electronic voting in the quantum era.
\end{abstract}
\bigskip

\maketitle

\section{Introduction}\label{intro}

Electronic voting systems have brought significant improvements in the speed, accessibility, and accuracy of elections. However, persistent security concerns threaten their reliability and public trust. Studies have documented vulnerabilities such as software tampering, weak cryptographic protections, and unauthorized access in widely used systems such as the Diebold AccuVote-TS \cite{kohno2004, balzarotti2009}. These weaknesses raise serious risks of vote manipulation, breaches of voter privacy, and challenges in verifying election outcomes.

The emergence of quantum computing further increases these threats by making today’s encryption breakable, enabling future decryption of stored data, and undermining trust in digital signatures and secure communication. Algorithms such as Shor’s algorithm could compromise widely adopted encryption methods, including RSA and elliptic curve cryptography \cite{prajapat2024, boyen2021}. As a result, there is growing interest in developing voting systems that are resilient against both classical and quantum attacks. Recent work emphasizes the role of Quantum Key Distribution (QKD) and post-quantum cryptography as key tools for securing digital elections in the coming era \cite{alshahrani2024}.

This paper proposes a quantum-secure voting framework that integrates QKD, Dual-Key Symmetric Encryption, and auditable receipt mechanisms. QKD ensures secure key exchange by making interception attempts detectable \cite{alleaume2014}, while homomorphic encryption allows votes to be tallied without revealing their contents \cite{katambo2019}, The proposed technique uses Dual-Key Symmetric Encryption to simulate the same with fewer computational resources. An auditable receipt system enables voters to verify that their votes were recorded correctly, enhancing transparency and trust.

Unlike previous approaches that mainly focus on blockchain or post-quantum cryptography \cite{mandal2024}, the proposed system combines multiple security layers into a unified architecture. To evaluate its performance, simulations were conducted using MQTT to replicate both classical and quantum communication scenarios. This work demonstrates a scalable and practical model for secure electronic voting in the quantum era, offering privacy, integrity, and verifiability without sacrificing usability.

The remainder of this paper is structured as follows. Section \ref{LR} presents a detailed review of the current state of voting technologies. Section \ref{Methodology} describes the proposed system architecture, and Methodology. Section \ref{sys} provides a comprehensive analysis of system performance, error handling, and potential vulnerabilities. Finally, Section \ref{con} summarizes the findings and outlines future directions for deployment in real-world elections.

\section{Literature Review }\label{LR}

\begin{figure}
    \centering
    \includegraphics[width=1\linewidth]{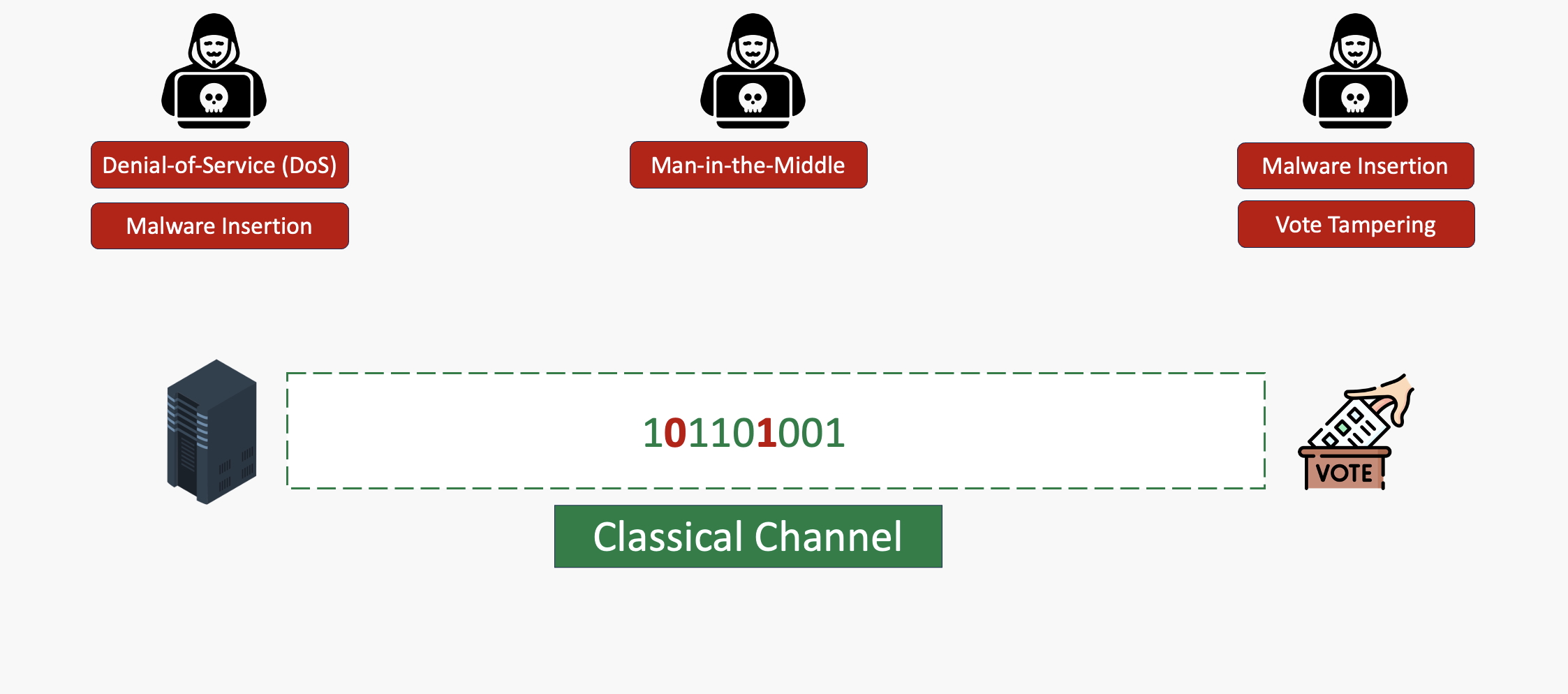}
    \caption{Attack Types on e-Voting Systems
}
    \label{fig:1}
\end{figure}

Electronic voting systems offer improvements in accessibility and efficiency but continue to face serious security vulnerabilities. Studies have shown risks such as vote tampering, malware insertion, denial-of-service attacks, and insider threats as shown in figure \ref{fig:1}, undermining the reliability and integrity of elections \cite{sundar2014, kohno2004, balzarotti2009, abba2017}. Systems like Diebold AccuVote-TS exhibited critical flaws, including poor cryptographic protections and insufficient access control mechanisms, making them susceptible to manipulation \cite{kohno2004, balzarotti2009}.

The emergence of quantum computing exacerbates these concerns by threatening classical cryptographic methods like RSA and elliptic curve cryptography \cite{prajapat2024, boyen2021}. Quantum Key Distribution (QKD) offers a promising solution by enabling secure key exchange that can detect eavesdropping \cite{sundar2014, alleaume2014}. Studies have explored the use of QKD alongside symmetric cryptography to enhance election security, while recognizing challenges like system complexity, cost, and the need for error correction techniques \cite{dervisevic2024, sharma2019, lovic2020, nurhadi2018, trizna2018}.

Several papers proposed quantum-enhanced election protocols. For instance, the author  of \cite{gao2021} developed a quantum election protocol achieving verifiability and receipt-freeness without pre-shared keys or a trusted third party. In \cite{zheng2021} the authors introduced a Quantum Digital Signature Verification Scheme suitable for electronic voting, improving efficiency and maintaining privacy through QKD integration. These advancements highlight the potential of quantum cryptographic methods but also underline limitations, such as reliance on trusted centers and the challenges of widespread deployment.

Homomorphic encryption also supports secure vote processing without decryption. Solutions such as lattice-based schemes \cite{wu2018}, Quantum Homomorphic Encryption \cite{chen2023}, and cloud-integrated architectures \cite{toapanta2020} aim to provide scalable and verifiable vote tallying.

Lastly, auditable receipt mechanisms are essential for improving voter trust and election transparency. Solutions like the Auditable Blockchain Voting System (ABVS) built on Hyperledger Fabric aim to create verifiable trails without sacrificing privacy \cite{mukherjee2020, pawlak2021}. Despite promising advancements, integrating QKD, Dual-Key Symmetric Encryption, and auditable systems remains necessary to fully address the evolving threats to electronic voting integrity in the quantum era.

\section{Methodology}\label{Methodology}

The proposed methodology integrates quantum and classical cryptographic techniques to ensure secure, private, and verifiable electronic voting. A symmetric key is first established between the voter and the election committee using the BB84 quantum key distribution (QKD) protocol, implemented with QASM 3 and enhanced through simulated noise handling. This key is used to encrypt the vote and voter ID via a bitwise XOR operation, enabling only authorized decryption. A dual-key symmetric encryption scheme is applied to preserve privacy during vote tallying, allowing aggregation without revealing individual identities. Verifiability is achieved through a receipt-based mechanism, where a hash of the encrypted data enables the voter to confirm successful vote registration. Finally, the MQTT protocol facilitates lightweight, secure communication across system components, supporting both classical and quantum data exchange.

\subsection{System Architecture} \label{arch}

\begin{figure}
    \centering
    \includegraphics[width=1\linewidth]{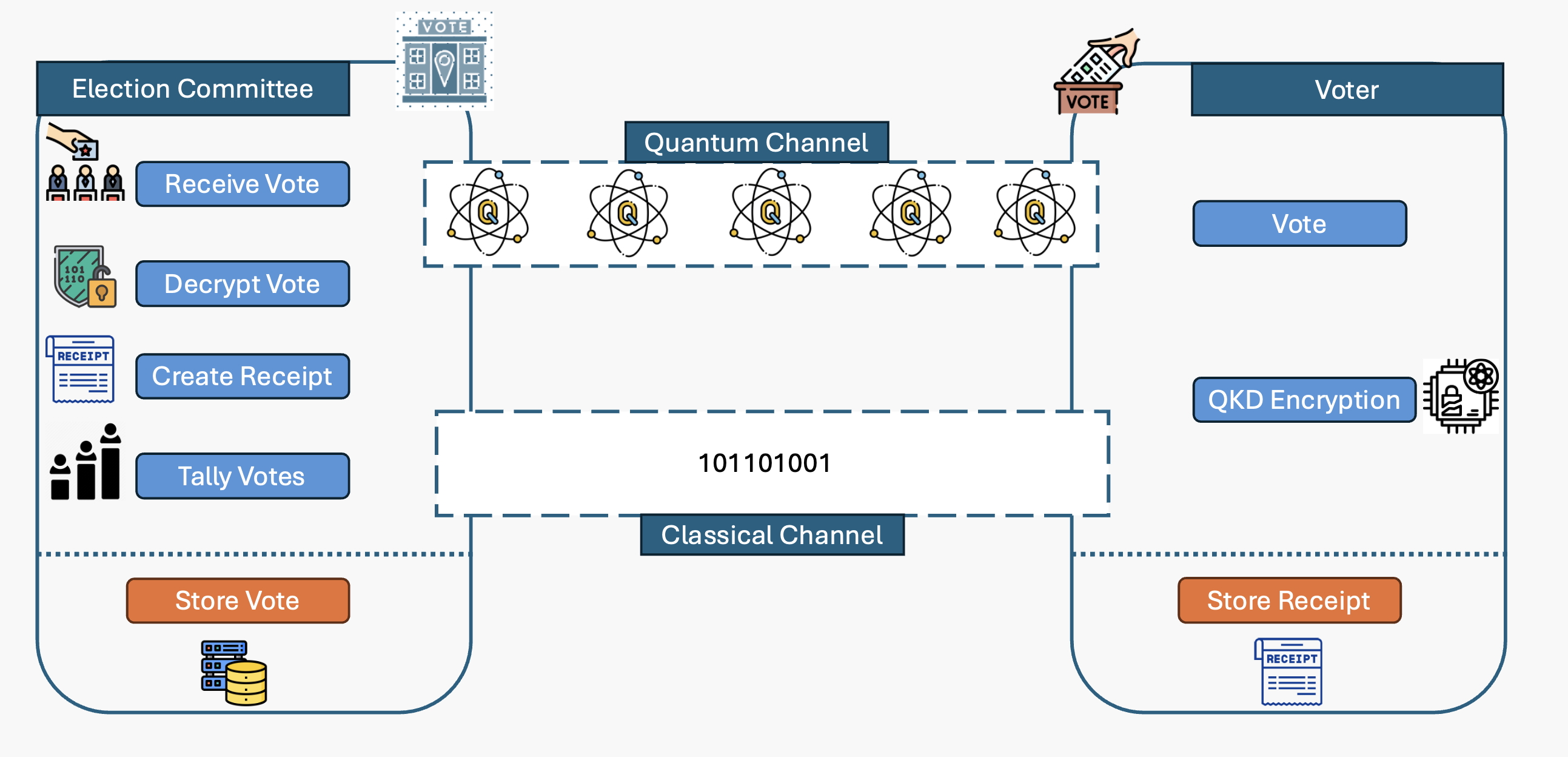}
    \caption{Proposed eVoting System}
    \label{fig:2}
\end{figure}

The architecture as shown in figure \ref{fig:2} is designed to securely transmit votes, protect voter anonymity during tallying, and enable transparent verification without compromising privacy.. Each voter first establishes a secure encryption key with the Election Committee server using QKD. The process involves encoding the key into qubits based on a randomly selected basis, such as rectilinear or diagonal, and transmitting it through a quantum channel. The corresponding bases are later shared over a classical channel to allow basis reconciliation. Leveraging the quantum no-cloning theorem \cite{buzek1996}, this setup ensures that any interception attempts are detectable, enhancing the security of key distribution. The BB84 protocol serves as the underlying QKD method to facilitate robust and practical secure communication \cite{elboukhari2009}.

Once the encryption key is successfully exchanged, the voter encrypts their vote and sends the ciphertext to the Election Committee server via the classical channel. Upon receiving the encrypted vote, the server decrypts it using the shared key and generates an auditable receipt confirming successful recording. This receipt is returned to the voter, allowing independent verification of vote inclusion without revealing ballot contents.

The proposed platform utilizes a Dual-Key Symmetric encryption to maintain voter privacy during tallying. This approach enables the aggregation of encrypted votes without exposing individual selections or requiring ballot decryption. The voter identity is stored separately from the voting data and accessed only in case of disputes, ensuring that privacy, integrity, and auditability are preserved throughout the election process. The following subsections detail the functionalities and interactions of each component. 

\subsection{Channel Simulation using Message Queuing Telemetry Transport (MQTT)}\label{MQTT}
\begin{figure}
    \centering
    \includegraphics[width=1\linewidth]{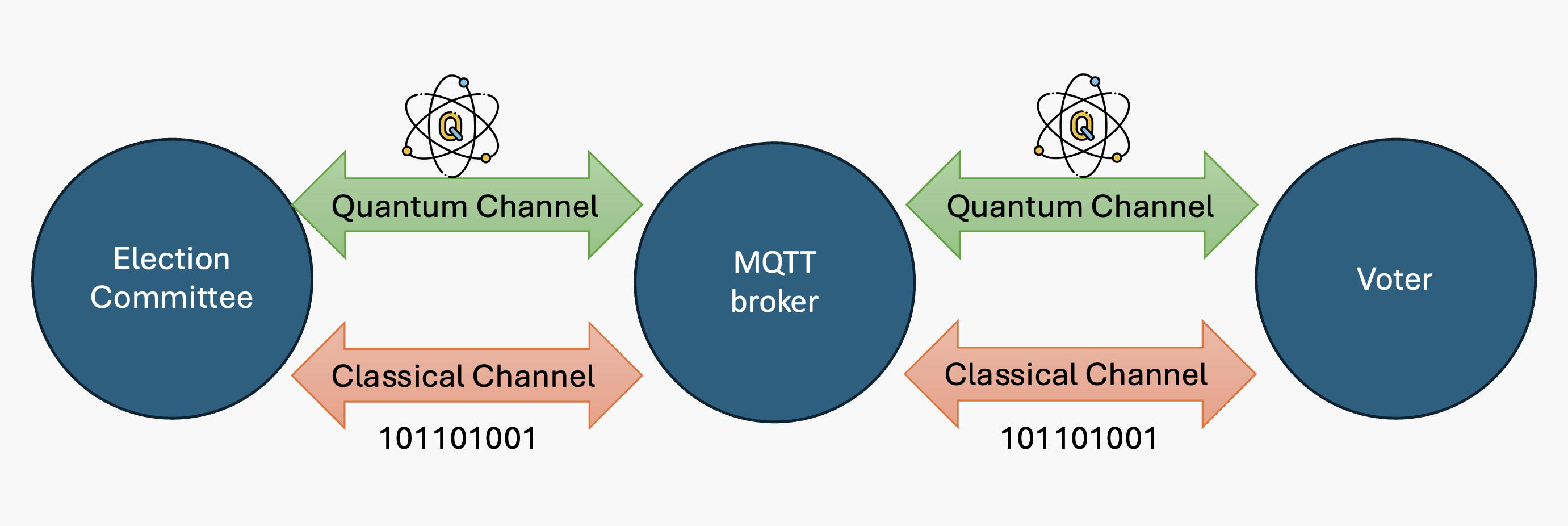}
    \caption{MQTT Broker}
    \label{fig:3}
\end{figure}

To simulate real-world network communication between voters and the Election Committee server, the system employs the Message Queuing Telemetry Transport protocol. MQTT is a lightweight, publish-subscribe messaging protocol designed for low-bandwidth, high-latency, or unreliable networks, making it well-suited for simulating both classical and quantum communication environments. In this setup, voters publish their encrypted votes and receipt verification messages to designated topics, while the server subscribes to these topics to receive and process incoming data as shown in figure \ref{fig:3}. The use of MQTT allows for efficient and reliable message delivery while minimizing communication overhead, thereby accurately modeling the transmission conditions that a real-world voting system may encounter. Furthermore, its lightweight nature facilitates rapid exchange of quantum keys and vote information without introducing significant network delays, enabling the evaluation of system performance under realistic conditions.

\subsection{Quantum Key Sharing via BB84}\label{BB84}

For this system, the BB84 protocol is chosen due to its established security and ease of implementation. The protocol facilitates secure key exchange between voters and the central server (voter and election committee). The system uses single-photon sources and polarization techniques to encode bits, ensuring that any eavesdropping attempts introduce detectable errors.

\begin{figure}
    \centering
    \includegraphics[width=1\linewidth]{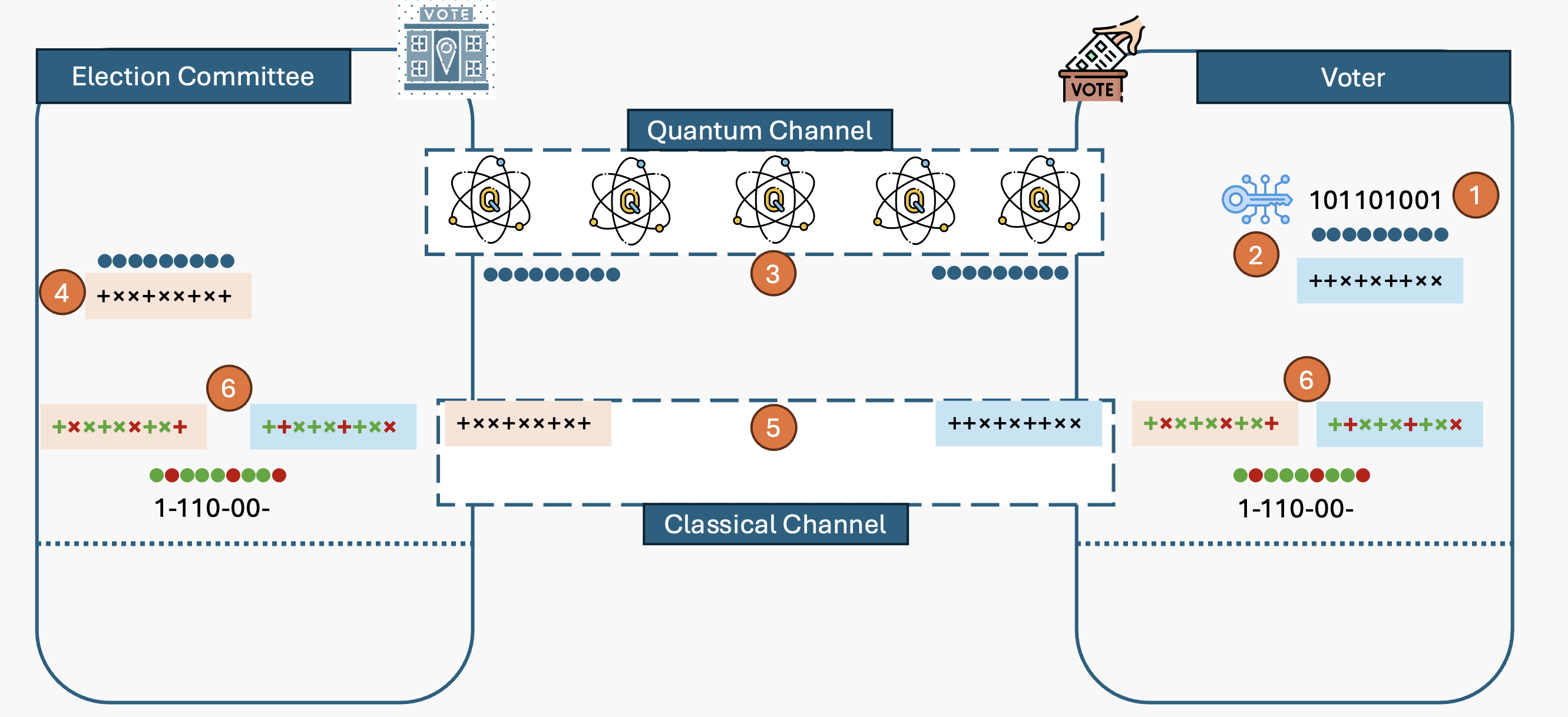}
    \caption{QKD Key Generation and Sharing Steps}
    \label{fig:4}
\end{figure}

Figure \ref{fig:4} illustrates the process of QKD BB84 protocol. The process is done following 6 steps. In Step 1 The voter initiates the process by generating a random key, which is a sequence of bits. In Step 2 this key is then encoded into a quantum state using randomly chosen bases: either the rectilinear basis (standard basis: $\ket{0} or \ket{1}$) or the diagonal basis (Hadamard basis: $\ket{+} or \ket{-}$). Each bit of the key is represented as a qubit using one of these bases, converting the classical key into its quantum form. Each QKD session creates new keys. These keys are used once to keep the XOR encryption secure. In Step 3 the voter then transmits the encoded qubits through the quantum channel to the Election Committee.

Upon receiving the qubits, In Step 4 the Election Committee measures them using a randomly selected basis (either rectilinear or diagonal) for each qubit. Since the Election Committee’s basis choice may not match the basis originally used by the voter, some measurements may yield incorrect results.

In Step 5 the voter and the Election Committee publicly announce the bases they used for each qubit over a classical communication channel. In Step 6 they compare their basis choices, and only the qubits measured using the same basis by both parties are retained. These matching bits (marked in green) are kept and form the shared key, while the bits from mismatched bases (marked in red) are discarded as garbage bits.

The voters and the Election Committee can reconstruct the same shared key with the matching bits. The voter then uses this key to encrypt the vote, and by the Election Committee to decrypt the information securely, ensuring that the shared key remains confidential and tamper-evident due to the properties of the quantum channel.

\subsubsection{QASM 3 for Quantum Information Sharing}

QASM 3 (Quantum Assembly Language 3) is an advanced, high-level assembly language designed to represent quantum circuits and operations for quantum computing \cite{cross2022}. In this system, QASM 3 enables precise encoding of quantum information, such as qubit states and measurement bases, in a standardized format. The voter uses QASM 3 to encode random bits and basis choices as qubits, which are then shared with the server over the simulated quantum channel. QASM 3’s flexibility and compatibility with quantum simulators like Qiskit Aer make it ideal for accurately defining and sharing quantum circuits, allowing the system to implement and simulate the BB84 quantum key distribution protocol effectively. By leveraging QASM 3, the system ensures that quantum information is transmitted in a universally interpretable format, supporting reliable and secure quantum communication.

\subsubsection{Simulating Quantum Errors and Noise Mitigation in QKD}

In the quantum voting system, an error rate is introduced to simulate quantum errors that occur naturally in real quantum environments due to noise and other quantum decoherence factors \cite{chen2022}. In the voter code, a random error rate between 0 and 0.2 is applied to each qubit transmitted over the quantum channel, with a chance that some qubits may flip states during transmission. This simulates the effect of quantum channel noise, where the integrity of qubit states can be compromised. The server mitigates these errors by only retaining bits measured in matching bases during the basis comparison phase, which effectively filters out mismatched and potentially erroneous bits. By ensuring that both voter and server rely on the agreed-upon shared key composed only of matching bits, the system minimizes the impact of quantum errors on the encryption and decryption processes, ensuring a robust and reliable quantum key distribution even under simulated noisy conditions.

Once a shared key is established between the voter and the server using the BB84 QKD protocol, it is employed to encrypt both the vote and the voter's ID through a simple bitwise XOR operation. Each bit of the vote and voter ID is XORed with the corresponding bit of the shared key, ensuring that only the intended server, which holds the matching key, can decrypt and recover the original information. Upon receiving the encrypted data, the server applies the XOR operation again to retrieve the original vote and voter ID, preserving privacy and ensuring the authenticity of voter registration.

\subsection{Privacy-Preserving Vote Tallying via Dual-Key Symmetric Encryption}\label{homomrphic}

\begin{figure}
    \centering
    \includegraphics[width=1\linewidth]{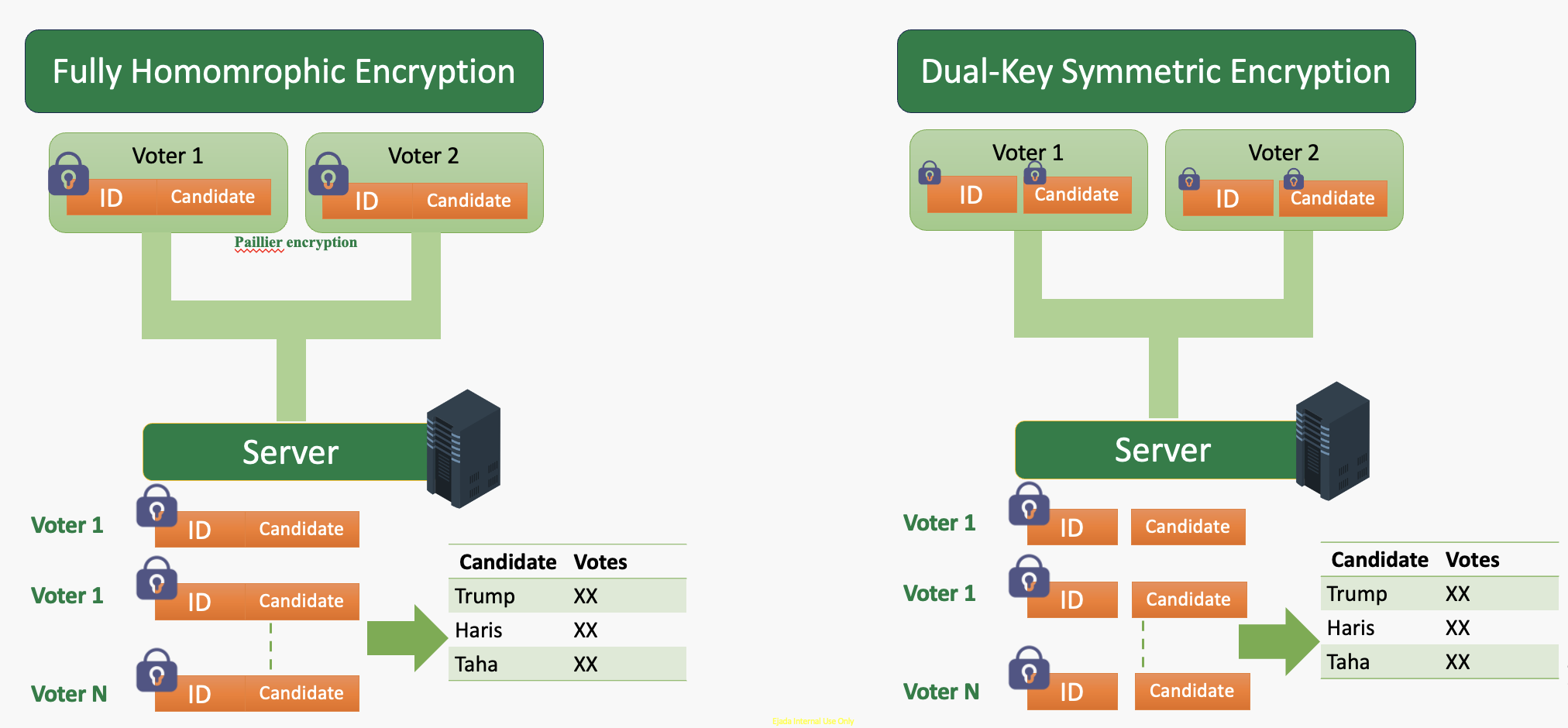}
    \caption{Difference between Fully Homomorphic and Dual-Key Symmetric Encryption}
    \label{fig:5}
\end{figure}

To preserve voter privacy during tallying, the proposed system uses a dual-key symmetric encryption strategy rather than a formal homomorphic encryption scheme. Each vote and its corresponding voter ID are encrypted separately using two different keys derived from QKD. The server receives only the key required to decrypt the vote, enabling it to tally results without accessing voter identities. This approach ensures privacy while allowing verification if needed: in case of audit or dispute, the voter can voluntarily submit the second key to decrypt their identity. This design simulates the effect of homomorphic tallying while maintaining low computational overhead and simplicity. \ref{fig:5} shows the difference between the Fully Homomorphic Encryption technique and the proposed Dual-Key symmetric encryption.

Although full homomorphic encryption, such as Paillier encryption \cite{paillier2005}, would allow operations on encrypted data without decryption, it introduces significant complexity and resource demands, especially in resource-constrained environments. In contrast, the Dual-Key Symmetric approach adopted here offers efficient, privacy-preserving aggregation with lower computational overhead. This method allows tallying without revealing identities, like homomorphic encryption, but with lower complexity, making it a practical choice for scalable, real-world voting systems where QKD-based symmetric encryption provides both security and efficiency.

\subsection{Receipt-Based Verification}\label{receipt}

The receipt hash is generated using SHA-256 by concatenating the encrypted vote and voter ID $H = \text{SHA256}(E_{\text{vote}} | E_{\text{id}})$. After receiving and processing the vote, the server sends the hash back to the voter, who recomputes it locally and compares it with the received value. A match confirms that the vote was received by the server and stored correctly, without requiring exposure of the original vote or identity.

\subsection{Protocol Summary}

The following pseudocode \ref{Algo} outlines the core steps of the proposed quantum-secure voting system using QKD and Dual-Key Symmetric Encryption:

\begin{algorithm}[H] \label{Algo}
\caption{Dual-Key Symmetric Voting Protocol}
\begin{algorithmic}[1]
\State \textbf{[Voter Side]}
\State Generate two QKD-derived symmetric keys: $K_{\text{vote}}$ and $K_{\text{id}}$
\State Encrypt vote: $E_{\text{vote}} = \text{XOR}(\text{vote}, K_{\text{vote}})$
\State Encrypt voter ID: $E_{\text{id}} = \text{XOR}(\text{ID}, K_{\text{id}})$
\State Compute local receipt hash: $H = \text{SHA256}(E_{\text{vote}} \, \| \, E_{\text{id}})$
\State Send $E_{\text{vote}}, E_{\text{id}}$ to Election Committee

\State \textbf{[Server Side]}
\State Decrypt $E_{\text{vote}}$ using $K_{\text{vote}}$
\State Generate and send back receipt hash: $H' = \text{SHA256}(E_{\text{vote}} \, \| \, E_{\text{id}})$
\State Do not decrypt $E_{\text{id}}$ unless audit is requested
\State Store $E_{\text{id}}$ securely for dispute resolution

\State \textbf{[Voter Side (Verification)]}
\State Recompute $H$ and compare with received $H'$ to verify vote was recorded
\end{algorithmic}
\end{algorithm}

\section{Results and Discussion} \label{sys}

The proposed quantum-secure voting system introduces a novel approach to achieving secure, privacy-preserving, and verifiable electronic voting. The system efficiently ensures voter privacy and vote integrity through Quantum Key Distribution, Dual-Key Symmetric encryption, and receipt validation. However, deploying such a system in practice introduces various challenges and considerations. This section discusses the system from multiple perspectives, including scalability, error detection and handling, security trade-offs, potential vulnerabilities, and recommendations for future improvements.

Initial testing showed that the system is capable of handling high volumes of encrypted vote processing efficiently. On a local machine equipped with an 8-core MacBook Air, the server processed each vote—including QKD simulation, encryption, and receipt validation—in approximately 0.0001 seconds. This equates to a theoretical capacity of up to 10 million votes in under 16 minutes. These results suggest that the core protocol is scalable under simulated conditions.

However, transitioning to a cloud-based infrastructure would likely enhance throughput and concurrency. Factors such as message queue handling, distributed key management, and network latency need to be addressed in large-scale deployments. Additionally, QKD was simulated using Qiskit with 10,000 shots per qubit, providing reliable results but lacking real-world imperfections such as photon loss and decoherence. Testing indicated that small keys (e.g., 2 bits) and large keys (e.g., 32 bits) were more error-prone, while a 4-bit key offered the most stable performance under noise simulation. This result emphasizes the importance of selecting optimal key sizes based on expected quantum channel conditions. Table \ref{tab:results_summary} shows the result summary, configuration used, and suggested enhancements.

\begin{table*}[t]
\centering
\caption{Summary of Experimental Results}
\begin{tabular}{|p{5cm}|p{10.5cm}|}
\hline
\textbf{Metric / Aspect} & \textbf{Value / Observation} \\
\hline
Vote Processing Time (per vote) & $\sim$0.0001 seconds \\
Estimated Throughput & $\sim$10 million votes in $\sim$16 minutes \\
Testing Environment & MacBook Air, 8-core CPU, Qiskit simulator \\
QKD Shot Count & 10,000 shots per measurement (stable performance) \\
Optimal Key Size for Stability & 4 bits \\
Unstable Key Sizes & 2 bits (too small), 32 bits (too large) \\
QKD Environment & Simulated; real hardware not used (quantum noise and latency not fully represented) \\
Encryption Scheme & Bitwise XOR using shared key from BB84 \\
Vote Tallying Method & Dual-Key Symmetric encryption (efficient, less resource-intensive than full HE)\\
MQTT Messaging & Functional in test; future work recommends migration to cloud-based solutions \\
\hline
\end{tabular}
\label{tab:results_summary}
\end{table*}

\subsection{Security and Privacy Trade-offs}

The system’s Dual-Key Symmetric encryption approach was selected to balance simplicity and privacy preservation. The Dual-Key Symmetric encryption approach used in this project is novel and meets privacy requirements with minimal complexity.

\subsection{Impact of Quantum Errors and Noise}

Quantum communication channels naturally face challenges like noise and signal loss, which can affect data accuracy. The system simulates quantum noise during the key exchange process to reflect this. It reduces errors by keeping only the qubits where both parties used the same measurement basis. Still, some mismatches occurred during testing. To address this, a re-send method could help recover from errors, and more advanced correction techniques may be needed in real-world use. The tests also showed that very short or very long keys caused more errors, while a 4-bit key gave the most reliable results. This suggests that choosing the correct key size is essential and should depend on the expected level of noise.

\subsection{Potential Vulnerabilities}

Despite the system's robust design, vulnerabilities in the classical channel persist, such as the possibility of man-in-the-middle attacks during basis announcements or message transmission. Additionally, timing attacks could exploit message delays to reveal patterns in data transmission. Enhancing classical channel security through encrypted, authenticated connections and implementing secure message validation methods would help mitigate these risks. As quantum cryptographic techniques evolve, future protocols may further strengthen classical communications within quantum-secure systems.

\subsection{Challenges of Real-World Deployment}

Scaling this system for real-world elections would require substantial infrastructure, particularly reliable quantum channels. QKD networks are currently geographically limited and often need dedicated optical or free-space links. Widespread adoption would entail significant investment and collaboration among telecommunications providers and regulatory bodies. Furthermore, ensuring high-speed quantum channels and expanding QKD networks would be critical for practical deployment, especially in areas where fiber-optic infrastructure is limited.

\subsection{Ethical and Societal Implications}

Quantum-secure voting systems have profound ethical and societal implications, particularly for public trust in digital elections. While this system enhances security, privacy, and transparency, it must also be user-friendly to gain public acceptance. Ensuring accessibility and reliability, especially for individuals with limited technical knowledge, is vital for fostering public trust. Additionally, the system’s ability to protect voter privacy and prevent fraud aligns with democratic values, potentially strengthening the legitimacy of digital elections and promoting broader acceptance of electronic voting systems.

\subsection{Comparison with Existing Voting Systems}

Compared to traditional electronic voting systems, this proposed system offers enhanced security by resisting quantum-based attacks that could potentially compromise classical encryption schemes. QKD provides a unique layer of security through tamper-evident key distribution. Dual-key symmetric encryption enables efficient aggregation of encrypted votes, a feature not typically feasible in conventional systems. This dual-key symmetric approach achieves the same functional objective as homomorphic encryption, which is privacy-preserving vote tallying, while requiring minimal computational resources, making it highly practical for real-world deployment and well-suited for integration with QKD-based secure key exchange. However, the complexity of implementing QKD and other quantum protocols presents challenges not encountered in traditional systems, highlighting the need for specialized infrastructure and quantum-compatible communication channels.

\section{Conclusion and Future Directions}\label{con}
This paper introduced a secure electronic voting system that combines quantum key distribution  with Dual-Key Symmetric encryption to protect both the privacy and integrity of the voting process. By using the BB84 protocol, the system ensures that encryption keys are securely shared in a tamper-evident way. At the same time, Dual-Key Symmetric encryption allows the server to tally encrypted votes without revealing individual choices, preserving voter anonymity. While the system performed well in simulation, there are still challenges to overcome, such as handling noise in quantum environments and ensuring the system can scale for large elections.

Looking ahead, several improvements could make the system more practical and efficient. One solution is to explore cloud-based communication tools, such as Azure IoT Hub or AWS SQS, instead of MQTT, which could improve reliability and handle more users at once. Finally, testing the system on real quantum hardware would give a better understanding of how it behaves under real-world conditions, including how well it manages noise and performance issues. These future steps would help bring quantum-secure voting systems closer to real-life applications, offering a strong foundation for secure and trustworthy digital elections.

\bibliographystyle{apalike}  
\bibliography{sn-bibliography}

\end{document}